\documentclass[10pt,journal]{IEEEtran}
\usepackage[pdftex]{graphicx}
\usepackage{algorithm}
\usepackage{color}
\usepackage{caption}
\usepackage{algpseudocode}
\newcommand{\RNum}[1]{\uppercase\expandafter{\romannumeral #1\relax}}
\definecolor{ForestGreen}{RGB}{162,52,0}
\usepackage{xcolor}
\usepackage{amsmath}
\usepackage{CJK}
\usepackage{url}
\usepackage{array}
\usepackage{ragged2e}
\usepackage{latexsym}
\usepackage[misc]{ifsym}

\ifCLASSOPTIONcompsoc
  \usepackage[nocompress]{cite}
\else
  \usepackage{cite}
\fi
\ifCLASSINFOpdf
\else
\fi
\ifCLASSOPTIONcompsoc
  \usepackage[caption=false,font=footnotesize,labelfont=sf,textfont=sf]{subfig}
\else
  \usepackage[caption=false,font=footnotesize]{subfig}
\fi

\begin{document}
\title{Security-Aware Virtual Network Embedding Algorithm based on Reinforcement Learning}

\author{Peiying Zhang,
Chao Wang,
Chunxiao Jiang\Letter,~\IEEEmembership{Senior~Member,~IEEE},
and Abderrahim Benslimane,~\IEEEmembership{Senior~Member,~IEEE}
\thanks{This work is partially supported by the Major Scientific and Technological Projects of CNPC (Grant No. ZD2019-183-006), and partially supported by ``the Fundamental Research Funds for the Central Universities" of China University of Petroleum (East China) (Grant No. 20CX05017A, 18CX02139A).}
\thanks{Peiying Zhang and Chao Wang are with the College of Computer Science and Technology, China University of Petroleum (East China), Qingdao 266580, China. E-mail: zhangpeiying@upc.edu.cn}
\thanks{Chunxiao Jiang is with the School of Information Science and Technology, Tsinghua University, Beijing 100084, China, with the Beijing National Research Center for Information Science and Technology, Beijing 100084, China.E-mail: jchx@tsinghua.edu.cn}
\thanks{Abderrahim~Benslimane is with the Department of Computer Science, University of Avignon, LIA CERI, Avignon, France. E-mail: abderrahim.benslimane@univ-avignon.fr}
}

\markboth{IEEE Transactions on Network Science and Engineering}%
{Shell \MakeLowercase{\textit{Peiying Zhang et al.}}: Bare Demo of IEEEtran.cls for IEEE Journals}

\maketitle

\begin{abstract}
Virtual network embedding (VNE) algorithm is always the key problem in network virtualization (NV) technology. At present, the research in this field still has the following problems. The traditional way to solve VNE problem is to use heuristic algorithm. However, this method relies on manual embedding rules, which does not accord with the actual situation of VNE. In addition, as the use of intelligent learning algorithm to solve the problem of VNE has become a trend, this method is gradually outdated. At the same time, there are some security problems in VNE. However, there is no intelligent algorithm to solve the security problem of VNE. For this reason, this paper proposes a security-aware VNE algorithm based on reinforcement learning (RL). In the training phase, we use a policy network as a learning agent and take the extracted attributes of the substrate nodes to form a feature matrix as input. The learning agent is trained in this environment to get the mapping probability of each substrate node. In the test phase, we map nodes according to the mapping probability and use the breadth-first strategy (BFS) to map links. For the security problem, we add security requirements level constraint for each virtual node and security level constraint for each substrate node. Virtual nodes can only be embedded on substrate nodes that are not lower than the level of security requirements. Experimental results show that the proposed algorithm is superior to other typical algorithms in terms of long-term average return, long-term revenue consumption ratio and virtual network request (VNR) acceptance rate.
\end{abstract}

\begin{IEEEkeywords}
Virtual network embedding, Network virtualization, Intelligent learning algorithm, Network security.
\end{IEEEkeywords}

\section{Introduction}\label{sec:introduction}

With the rapid development of social economy, the number of network end users shows a blowout growth trend, which is expected to reach 40 billion in the near future \cite{B1,D1}. A large number of terminal resource requests bring great pressure to the underlying network. Because the Internet only provides "best effort" resource delivery, it cannot allocate the underlying resources reasonably and efficiently, so it gradually becomes rigid \cite{C1}. In recent years, NV has gradually come into people's vision. It is considered to be an efficient and dynamic network framework for managing network resources. The virtual network consists of several virtual nodes (such as virtual router and virtual server), which are connected by several virtual links. The problem of VNE is to map virtual network to shared substrate network and provide sufficient computing and bandwidth resources for requests \cite{B2,B3}. Radio network resource management faces severe challenges, including storage, spectrum, computing resource allocation, and joint allocation of multiple resources \cite{jcx1,jcx2}. With the rapid development of communication networks, the integrated space-ground network has also become a key research object \cite{jcx3}.

It cannot be ignored that NV brings flexibility to network architecture, but also brings some new security problems. In the NV environment, some VNRs require high security. For example, in recent years, online payment and online shopping, which are closely related to money, have become more and more popular. Some VNRs have relatively low security requirements, such as online chat and online video \cite{D2}. Because a large number of terminal devices need to request network resources, when the network is in a "busy" state, it is easy to ignore the security issues. At this time, the network may be attacked by some malicious software or cause the leakage of important information. Therefore, it is necessary to consider the security problem in virtual network mapping \cite{B4,C4}.

The problem of VNE has been proven to be NP-hard. Most of the traditional solutions are heuristic algorithms. By making a series of rules and constraints, they embed every VNR manually. In addition, most heuristic algorithms divide the VNE process into two stages: node mapping and link mapping \cite{B5}. However, the relationship between these two stages is not fully considered. In this way, the mapping results may fall into the local optimal solution. Due to the rich network features, resource constraints and location constraints are usually used to represent nodes, bandwidth constraints and delay constraints are used to represent links, which is not perfect.

In recent years, with the rise of cloud computing, artificial intelligence, machine learning (ML) and other emerging fields, using intelligent learning algorithm to solve the problem of VNE has become a trend \cite{D3,D5}. ML algorithms process a large amount of data collected over a period of time, automatically learn the required information from the data, then classify or predict \cite{B6, B7, C2}. As an excellent representative of ML, RL can be used to solve the problem of VNE. We incorporate the RL algorithm with the VNE algorithm with security awareness. The RL agent is trained and tested by reasonably extracting the attributes of the substrate nodes. Considering the security requirements of VNRs, we set the security level attribute for substrate nodes and the security requirement level attribute for virtual nodes. Finally, the experimental results show that our algorithm has achieved good results.

As far as we know, there is no research on the combination of security of VNE and intelligent learning algorithm to address the VNE problem. The main contributions of this article are as follows.

(1) This paper proposes a security aware VNE algorithm based on RL. Under the condition that the basic virtual network is embedded (the computational resource constraint and link bandwidth constraint of the node), the security requirement level constraint is bounded to each virtual node. Security level constraint is bounded to each substrate node. Virtual nodes can only be embedded in substrate nodes no lower than the security requirement level. This can ensure the security of VNE algorithm.

(2) We mainly apply RL algorithm to node embedding stage. Specifically, the whole algorithm is divided into training stage and testing stage. In the training phase, we use the policy network to train the learning agent. We take the five features of the substrate nodes as the input of the policy network. The result is to deduce the probability of each substrate node and map the virtual nodes according to the probability. The shortest path algorithm based on BFS is used for link mapping. In the test phase, the training results are directly utilized to complete the embedding of VNR.

(3) We compare our algorithm with other representative algorithms in terms of long-term average revenue, long-term revenue consumption ratio and VNR acceptance rate. The experimental results show that the algorithm based on RL is better than other algorithms. The security level constraint can also be applied to the VNE problem, therefore, it has some practical significance.

The rest of this article is organized as follows. The second part describes the related work of the VNE algorithms. The third part models security aware virtual network embedding process and proposes evaluation metrics. The fourth part introduces the RL algorithm based on policy network in detail. The fifth part introduces and analyzes the simulation experiments and experimental results. Finally, the whole article is summarized and prospected.

\section{Related Work}

\subsection{Virtual Network Embedding Related Algorithms}
The traditional heuristic algorithm can be divided into single-stage mapping algorithm and two-stage mapping algorithm according to the application of the model in the mapping stage. The difference is whether the node map and the link map are mapped simultaneously. In reference \cite{A3}, a mixed integer regularization algorithm is proposed. In this algorithm, node mapping process and link mapping process are considered to be a whole. Two mapping algorithms of certainty and randomness are obtained by relaxing integer constraints. After the virtual node mapping process, the multi commodity flow algorithm is used to complete the link mapping process. Vhub linear programming method is adopted in reference \cite{A4}. The VNE problem is treated as mixed integer programming problem using the p-hub median method. The best location of VNE can be determined after the location problem of hub is solved. Reference \cite{A5} proposes a two-stage VNE algorithm for path separation and migration. The algorithm fully measures the capability of the underlying network and fully considers the embeddability of the virtual network. Then the strategy of path segmentation and path migration is proposed. It effectively utilizes the substrate bandwidth and improves the robustness of the mapping strategy.

The security aware VNE algorithm based on RL also divides the VNE problem into two stages: node mapping and link mapping. Therefore, it belongs to two-stage VNE algorithm. Different from the above algorithm, the algorithm proposed in this paper does not use heuristic method to solve the problem of VNE. With the rapid development of intelligent learning algorithm in recent years, RL algorithm has been proven to be an efficient way to solve practical problems. Therefore, this paper uses RL to solve the problem of VNE.

\subsection{Security-Aware Virtual Network Embedding Algorithms}
At present, some researches have discussed the security problems of VNE. Liu et al \cite{A1} proposed a security VNE algorithm based on multi-attribute evaluation and path optimization. He modeled the mapping process of security virtual network as a multi-objective mixed integer linear programming model, and completed the embedding of virtual network by establishing node mapping function and link mapping function. However, his algorithm does not fully consider the security performance of each virtual node and physical node, but gives the security level to the entire VNR, which is not rigorous. Gong et al \cite{A2} proposed a trust aware security VNE algorithm. He introduced trust relationship and trust degree into virtual network resource allocation, and quantitatively analyzed the security problems in NV environment. The method of approaching ideal ordering is used to rank the importance of multi-attribute nodes. However, the algorithm makes the rules of VNE manually, which is not in line with the reality.

The security aware VNE algorithm based on RL and the above two algorithms also pay attention to the security of VNE. The main solution is to add security attributes to the virtual network and the substrate network. Different from the above algorithm, we set the security requirement level for each virtual node and set the security level for each substrate node, rather than setting the security attributes for the entire VNR. In addition, the above algorithm uses heuristic method to solve the problem of VNE. We adopt RL strategy to solve this practical problem. Practice shows that RL algorithm is a more efficient method.

\subsection{Machine Learning based Virtual Network Embedding Algorithms}
The above heuristic methods to solve the VNE problem cannot fully reflect the real situation of the network in reality. Most of them are based on artificial rules and cannot automatically optimize network parameters, which may lead to local optimization of embedding results. At present, a large number of scholars have used ML algorithms to solve the problem of VNE. Reference \cite{A6} proposed monte carlo search tree algorithm. The algorithm considers the node mapping process as a Markov decision process (MDP). When the VNR arrives, the monte carlo search tree is used to embed the node. Then the shortest path algorithm or multi-commodity flow algorithm is used for link mapping. Reference \cite{A7} introduced the neural network algorithm into the VNE problem. This algorithm proposes an autonomous system based on artificial neural network to improve the mapping efficiency of virtual network. In reference \cite{A8}, Q-learning algorithm in reinforcement learning is used to solve the VNE problem by using the optimal mapping mechanism of reward mechanism learning. Reference \cite{A9} proposes to use the Policy Gradient algorithm in the RL algorithm and gradually learn the optimal mapping mechanism by using the RL agent. The algorithm applies the Policy Gradient method to the VNE domain and mainly to the node embedding stage. This model explores how to strike a balance between exploring better solutions and developing existing models.

Our algorithm also uses a policy network as a learning agent to derive the probability of each substrate node. It should be noted that our work is different from the above research. It is the first time to combine the security attribute with the intelligent learning algorithm to study the problem of VNE. Secondly, we extract five important node attributes to train learning agents, including the key security attributes. More importantly, the overall performance of the algorithm is kept at a good level when the node attributes are extracted reasonably. More feature extraction allows the agent to learn more about the substrate network, making the VNE algorithm more practical. In addition, the above algorithms do not pay attention to the impact of security factors on the VNE algorithm. Our algorithm adds security attributes to nodes, which can well meet the security requirements of VNE.

\section{Network Models and Evaluation Indicators}

\subsection{Network Models}

\begin{figure}[!htp]
\includegraphics[width=1.0\columnwidth]{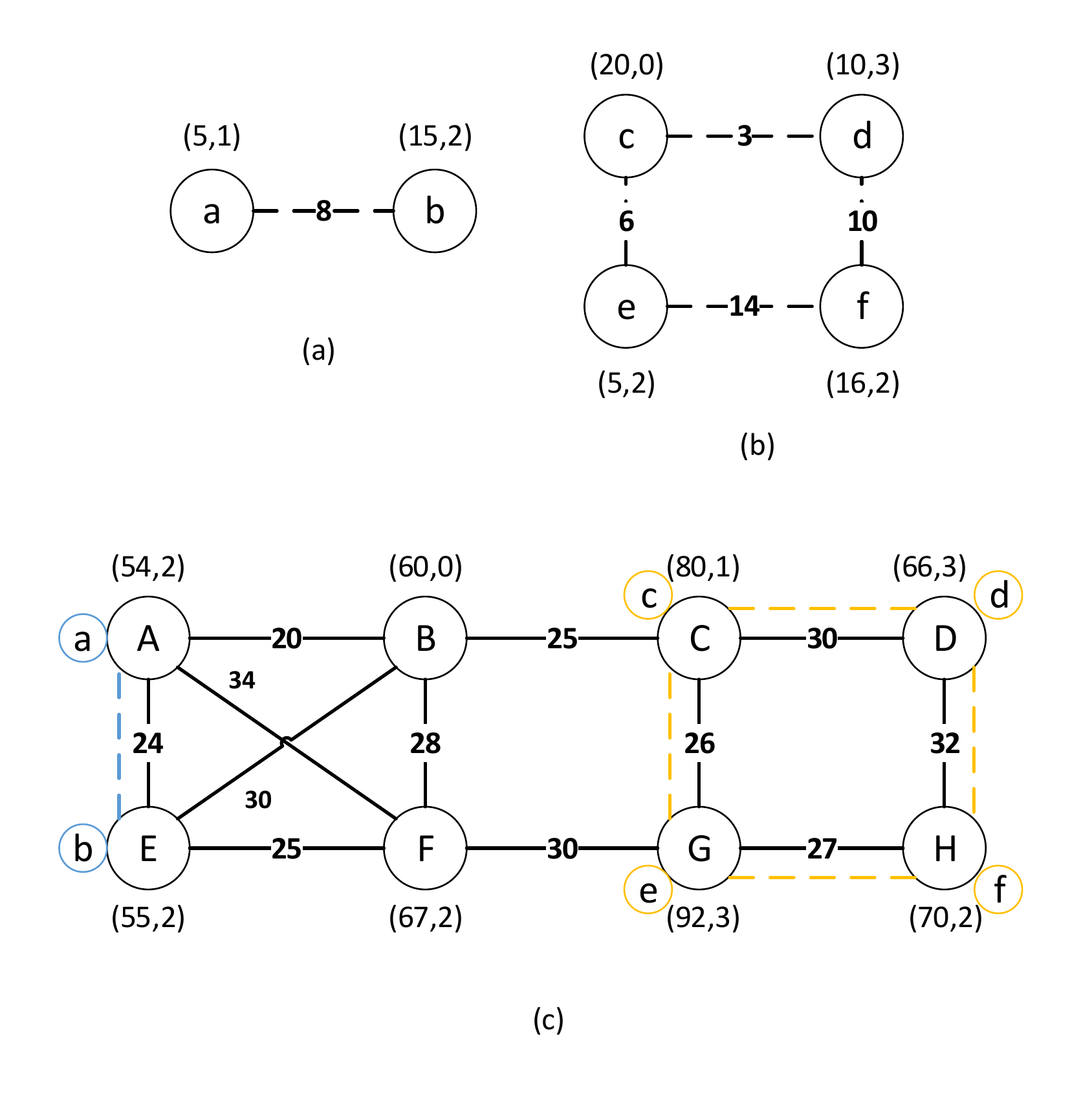}
\caption{Schematic diagram of virtual network embedding substrate network.}
\label{fig_2}
\end{figure}

The substrate network can be modeled as an undirected weighted graph $G^S=\{N^S,L^S\}$. Where $N^S$ represents the set of all substrate nodes and $L^S$ represents the set of all substrate links. Substrate node $n^s \in N^S$, whose attributes are represented by computing power $CPU(n^s)$ and security level $sl(n^s)$. The security level of substrate node is an important embodiment of substrate network security. The higher the level, the more secure the proof maps to the substrate node, and the less vulnerable it is to security issues. Substrate link $l^s \in L^S$, whose attributes are expressed as bandwidth capability $BW(l^s)$. (c) in Figure \ref{fig_2} represents a substrate network.

The same undirected weighted graph $G^V=\{N^V,L^V\}$ is used to model the virtual network. Where $N^V$ represents the set of all virtual nodes and $L^V$ represents the set of all virtual links. Virtual node $n^v \in N^V$, whose attributes are represented by calculated resource requirement $CPU(n^v)$ and security requirement level $sr(n^v)$. The security requirement level of virtual node represents the security requirement of VNR. In order to ensure the security of VNRs, virtual nodes can only be mapped to substrate nodes no less than their security requirements. Virtual link $l^v \in L^V$, whose attributes are represented by bandwidth resource requirement $BW(l^v)$. The (a) and (b) in Figure \ref{fig_2} represent two virtual networks.

We summarize all the symbols in TABLE \uppercase\expandafter{\romannumeral1}.

\begin{table}
\centering
\caption{Symbol summary}
\renewcommand\arraystretch{1.5}
\begin{tabular}{|p{15mm}|p{65mm}|}
\hline
Symbol & Description  \\
\hline
$G^S$ & substrate network  \\
\hline
$G^V$ & virtual network \\
\hline
$N^S$ & set of all substrate nodes \\
\hline
$L^S$ & set of all substrate links \\
\hline
$CPU(n^s)$ & the computational power of a substrate node \\
\hline
$BW(l^s)$ & The amount of bandwidth resources of a substrate link \\
\hline
$sl(n^s)$ & substrate node security level  \\
\hline
$N^V$ & set of all virtual nodes \\
\hline
$L^V$ & set of all virtual links \\
\hline
$CPU(n^v)$ & compute resource requirements for a virtual node \\
\hline
$BW(l^v)$ & bandwidth resource requirements for a virtual link \\
\hline
$sr(n^v)$ & security level requirements for a virtual node \\
\hline
\end{tabular}
\end{table}

The VNE problem can be expressed as $G^V(N^V,L^V) \to G^S(N^{S_i},L^{S_i})$, where $N^{S_i} \in N^S$, $L^{S_i} \in L^S$. The above process needs to meet the following constraints:

\begin{equation}
\begin{aligned}
\sum_{i=1}^{|N^V|}(n^v,n_i^v) \ne 0.
\end{aligned}
\end{equation}

\begin{equation}
\begin{aligned}
\sum_{j=1}^{|N^S|}(n^s,n_j^s) \ne 0.
\end{aligned}
\end{equation}

Formulas (1) and (2) indicate that neither a virtual node $n^v$ nor a substrate node $n^s$ exist independently. There must be other nodes connected to it.

\begin{equation}
\begin{aligned}
\sum_{j=1}^{|N^S|}(n_i^v \to n_j^s)=1.
\end{aligned}
\end{equation}

Formula (3) indicates that a virtual node $n_i^v$ can only be embedded into a physical node $n_j^s$, where $n_i^v \in N^V$, $n_j^s \in N^S$.

\begin{equation}
\begin{aligned}
{\forall}n_i^v \in VNR_k, \sum_{n_j^s \in N^S}^{|N^s|}n_{ij}^{vs} \leq 1.
\end{aligned}
\end{equation}

Formula (4) indicates that the virtual node $n_i^v$ in the same VNR cannot be mapped to the same substrate node $n_j^s$.

\begin{equation}
\begin{aligned}
\sum_{j=1}^{|L^S|}(l_i^v \to l_j^s) \geq 1.
\end{aligned}
\end{equation}

Formula (5) indicates that a virtual link $l_i^v$ can be embedded into one or more substrate links $l_j^s$, where $l_i^v \in L^V$, $l_j^s \in L^S$.

\begin{equation}
\begin{aligned}
n_{ij}^{vs}CPU(n_i^v) \leq n_{ij}^{vs}CPU(n_j^s).
\end{aligned}
\end{equation}

In formula (6), $CPU(n_i^v)$ represents the computing resource demand of the virtual node $n_i^v$. $CPU(n_j^s)$ represents the computing resource available from the substrate node $n_j^s$.

\begin{equation}
\begin{aligned}
l_{ij}^{vs}BW(l_i^v) \leq l_{ij}^{vs}BW(l_j^s).
\end{aligned}
\end{equation}

In formula (7), $BW(l_i^v)$ represents the bandwidth demand of virtual link $l_i^v$. $BW(l_j^s)$ represents the bandwidth resources available for the substrate link $l_j^s$.

\begin{equation}
\begin{aligned}
sl(n^s) \geq sr(n^v).
\end{aligned}
\end{equation}

In formula (8), $sl(n^s)$ represents the security level of the substrate node. $sr(n^v)$ represents the level of security requirements for the virtual node.

Figure \ref{fig_2} shows two virtual networks embedded in a substrate network. For two virtual networks, the first number in parentheses next to the node represents the calculated resource requirements of the node. The second number represents the level of security requirements for the node. The number on a virtual link represents the bandwidth requirements for that link. For the substrate network, the first number in parentheses next to the node represents the computing resources that the node can provide. The second number represents the security level of the node. The number on a substrate link represents the amount of bandwidth resources that the link can provide. Figure \ref{fig_2} shows the successful embedding of two VNRs into the base network. The corresponding relation of specific nodes is: $a \to A$, $b \to E$, $c \to C$, $d \to D$, $e \to G$, $f \to H$.

\subsection{Evaluation Indicators}

We evaluate the security aware VNE algorithm based on RL from three aspects: long-term average revenue, long-term revenue consumption ratio and VNR acceptance rate.

The VNE revenue is represented by $Re(G^V,t,t_p)$, where $t_p$ represents the duration of the VNR to arrive. Specifically, it is calculated based on node computing resource consumption $CPU(n^v)$ and link bandwidth resource consumption $BW(l^v)$. The expression method is shown in formula (9):

\begin{equation}
\begin{aligned}
Re(G^V,t,t_p)=t_p \cdot [\sum_{n^v \in N^V}CPU(n^v)+\sum_{l^v \in L^V}BW(l^v)].
\end{aligned}
\end{equation}

The VNE consumption is expressed as $Co(G^V,t,t_p)$. Specifically, it is calculated according to the calculated resource consumption $CPU(n^v)$ of the node and the total bandwidth resource consumption $BW(l_v^s)$ of the embedded multiple substrate links. As shown in formula (10):

\begin{equation}
\begin{aligned}
Co(G^V,t,t_p)=t_p \cdot [\sum_{n^v \in N^V}CPU(n^v)+\sum_{l^v \in L^V}\sum_{l^s \in L^S}BW(l^s_v)].
\end{aligned}
\end{equation}

The long-term average revenue is shown in formula (11):

\begin{equation}
\begin{aligned}
Avg\_Re=\lim_{Time \to \infty} \frac{\sum_{t=0}^{Time}R(G^V,t,t_p)}{Time},
\end{aligned}
\end{equation}
where $Time$ is the elapsed time.

Long-term revenue consumption is shown in formula (12):

\begin{equation}
\begin{aligned}
RC=\lim_{Time \to \infty} \frac{\sum_{t=0}^{Time}Re(G^V,t,t_p)}{\sum_{t=0}^{Time}Co(G^V,t,t_p)}.
\end{aligned}
\end{equation}

The VNR acceptance rate can be expressed as follows:

\begin{equation}
\begin{aligned}
Acp=\lim_{Time \to \infty} \frac{\sum_{t=0}^{Time}Accept(G^V,t,t_p)}{\sum_{t=0}^{Time}Arrive(G^V,t,t_p)},
\end{aligned}
\end{equation}
where $Accept(G^V,t,t_p)$ represents the number of successful VNRs mapped in the time range $t_p$. $Arrive(G^V,t,t_p)$ represents the total number of VNRs that arrive within the time range $t_p$.

\section{Introduction of Reinforcement Learning Algorithm Based on Policy Network}

\subsection{Extraction of Substrate Node Attributes}

We need to train RL agents in a substrate network as close to reality as possible, so we need to create a more "real" environment for agents. Because there are many properties of the substrate nodes, it will increase the computational complexity to represent them. We extract the following five attributes to represent the substrate nodes as input to the policy network.

(1) Computing capacity (CPU) : Computing capacity is one of the most important attributes to represent a node. The stronger the computing power of the node, the greater the probability that the substrate node receives the virtual node. The CPU can be represented as follows:

\begin{equation}
\begin{aligned}
CPU(n^s)_r=CPU(n^s)-\sum_{n^v \to n^s}CPU(n^s),
\end{aligned}
\end{equation}
where $CPU(n^s)_r$ represents the remaining computing power of the substrate node. $CPU(n^s)$ represents the initial computing power of the substrate node. $\sum_{n^v \to n^s}CPU(n^s)$ represents the sum of computational resources consumed by all VNRs for virtual nodes embedded in $n^s$.

(2) Degree (DEG) : The number of substrate links connected to the substrate node is called degree. The greater the degree of a node, the more nodes it is connected to. DEG can be expressed as:

\begin{equation}
\begin{aligned}
DEG(n^s)=\sum_{n_i^s \in N^S}Link(n^s,n_i^s),
\end{aligned}
\end{equation}
if $n^s$ is connected to $n_i^s$, $Link(n^s,n_i^s)=1$; if not, it is 0.

(3) Sum of bandwidth (SUM\_BW) : The sum of the bandwidth of all the links connected to a substrate node. The larger the node bandwidth and the larger the virtual node that is mapped to the substrate node will have more link options, resulting in a better mapping effect. SUM\_BW can be expressed as:

\begin{equation}
\begin{aligned}
SUM\_BW=\sum_{l^s \in L_n^S}BW(l^s),
\end{aligned}
\end{equation}
where $L_n^S$ represents the substrate link connected to node $n^s$. $l^s$ is one of $L_n^S$.

(4) Average distance from mapped node to this node (AVG\_DIS) : This property is considered for the link-mapping phase. The above attributes take into account the local importance of the node, which takes into account the global importance of the node. This property depicts the average distance to the mapped node, so the smaller the attribute, the greater the probability of the node being mapped. Finally, the shortest path algorithm based on BFS is used to map the link. AVG\_DIS can be expressed as:

\begin{equation}
\begin{aligned}
AVG\_DIS=\frac{\sum_{N_v^s \in N^S}DIS(n^s,n_v^s)}{count+1},
\end{aligned}
\end{equation}
where $DIS(n^s,n_v^s)$ represents the distance from $n^s$ to the mapped node. $count$ is the number of nodes that have been mapped, plus 1 is to prevent the denominator from being 0.

(5) Security level (SL) : The higher the security level of the substrate node, the safer the mapping to the node. Virtual nodes can only be mapped to substrate nodes with a higher level of security requirements.

We characterize the above properties of the i-th substrate node as a 5-dimensional vector, as follows:

\begin{equation}
\begin{aligned}
v_i=(CPU(n_i^s),DEG(n_i^s),SUM\_BW(n_i^s)\\,AVG\_DST(n_i^s),SL(n_i^s)).
\end{aligned}
\end{equation}

The attribute vectors of all the substrate nodes are put into a feature matrix $FM$, which is taken as the input of the policy network.

\begin{equation}
\begin{aligned}
FM=(v_1,v_2...v_n)^T.
\end{aligned}
\end{equation}

The feature matrix is expressed as follows:

\begin{equation}
\begin{aligned}
\begin{bmatrix}
  \begin{smallmatrix}
    CPU(n_1^s) & DEG(n_1^s) & SUM\_BW(n_1^s) & AVG\_DIS(n_1^s) & SL(n_1^s)\\
    CPU(n_2^s) & DEG(n_2^s) & SUM\_BW(n_2^s) & AVG\_DIS(n_2^s) & SL(n_2^s)\\
    ...        & ...        & ...            &...              & ...      \\
    CPU(n_k^s) & DEG(n_k^s) & SUM\_BW(n_k^s) & AVG\_DIS(n_k^s) & SL(n_k^s)\\
  \end{smallmatrix}
\end{bmatrix}
.
\end{aligned}
\end{equation}

\subsection{Policy Network}
\begin{figure}[!htp]
\includegraphics[width=1.0\columnwidth]{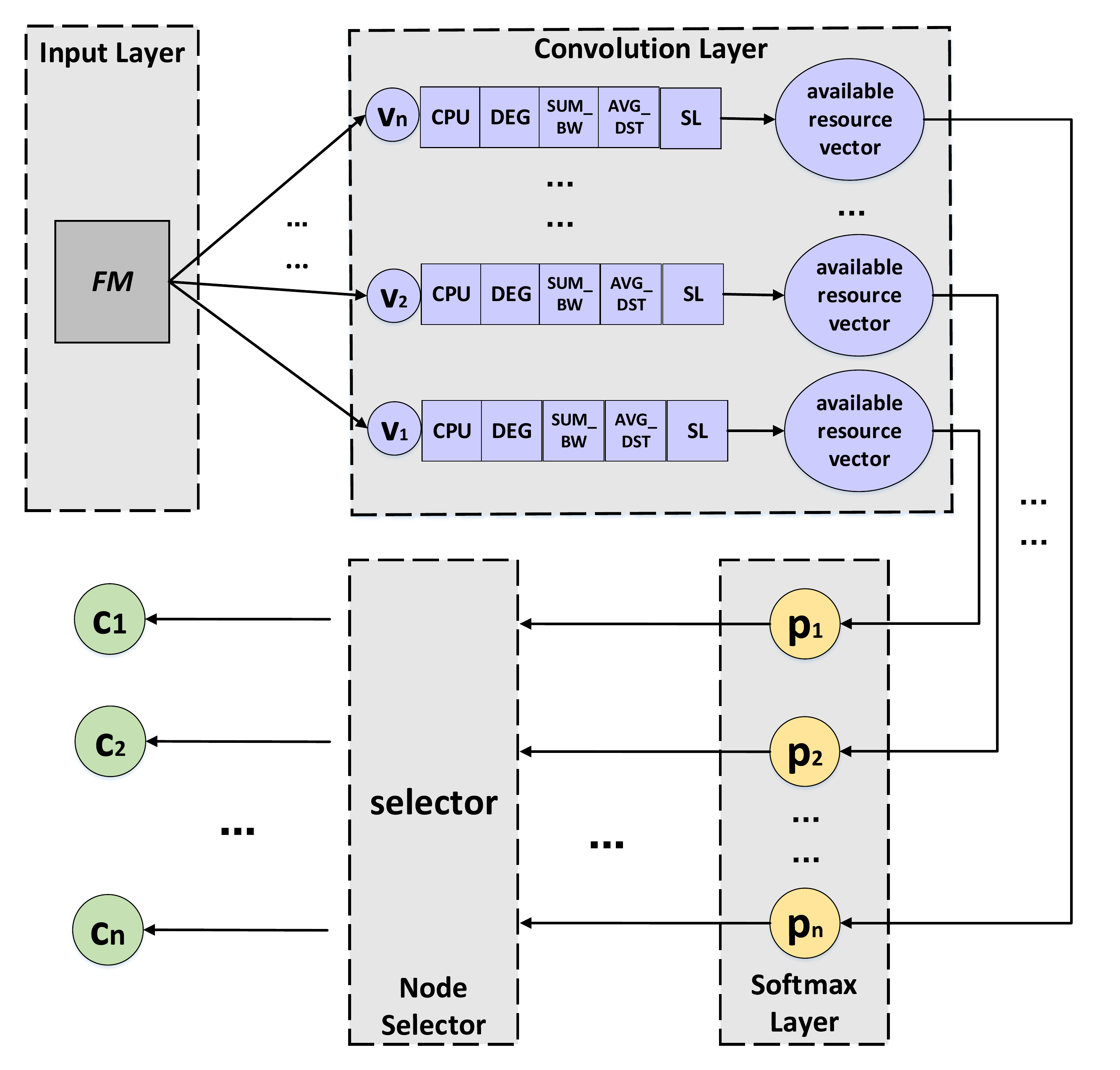}
\caption{Policy network.}
\label{fig_3}
\end{figure}

We use a policy network as a learning agent, which is essentially a convolutional neural network commonly used in ML \cite{B8,D4}. Taking the policy matrix as input, the mapping probability of each substrate node is output by training the learning agent. The greater the probability, the more likely the substrate node is to be mapped. The policy network consists of an input layer, a convolution layer, a softmax layer and a node selector, as shown in Figure \ref{fig_3}.

As the input of the policy network, the feature matrix is transferred from the input layer to the convolution layer. The main function of convolution layer is to convolute the feature matrix. Convolution operation originally refers to the operation of generating the third function from two functions. Here, the feature vector of each node can be obtained after convolution of the feature matrix. We call it the available resource vector, which is specifically expressed as:

\begin{equation}
\begin{aligned}
arv_i= \omega \cdot v_i + d,
\end{aligned}
\end{equation}
where $arv_i$ is the $i$-th output of the convolution layer, $\omega$ is the weight vector of the convolution kernel and $d$ is the deviation.

The vector is then transferred to the softmax layer and the softmax function of logistic regression is used to generate a probability for each node \cite{A9}. The higher the probability, the more likely the virtual node is to map to that node. Some substrate nodes may not be able to map part of the virtual nodes due to insufficient computing power or security level, so the probability of this part of substrate nodes being mapped cannot be deduced. We add a node selector to select a group of candidate nodes with enough computing power and security level.

\subsection{Training and Testing}
\begin{figure}[!htp]
\includegraphics[width=1.0\columnwidth]{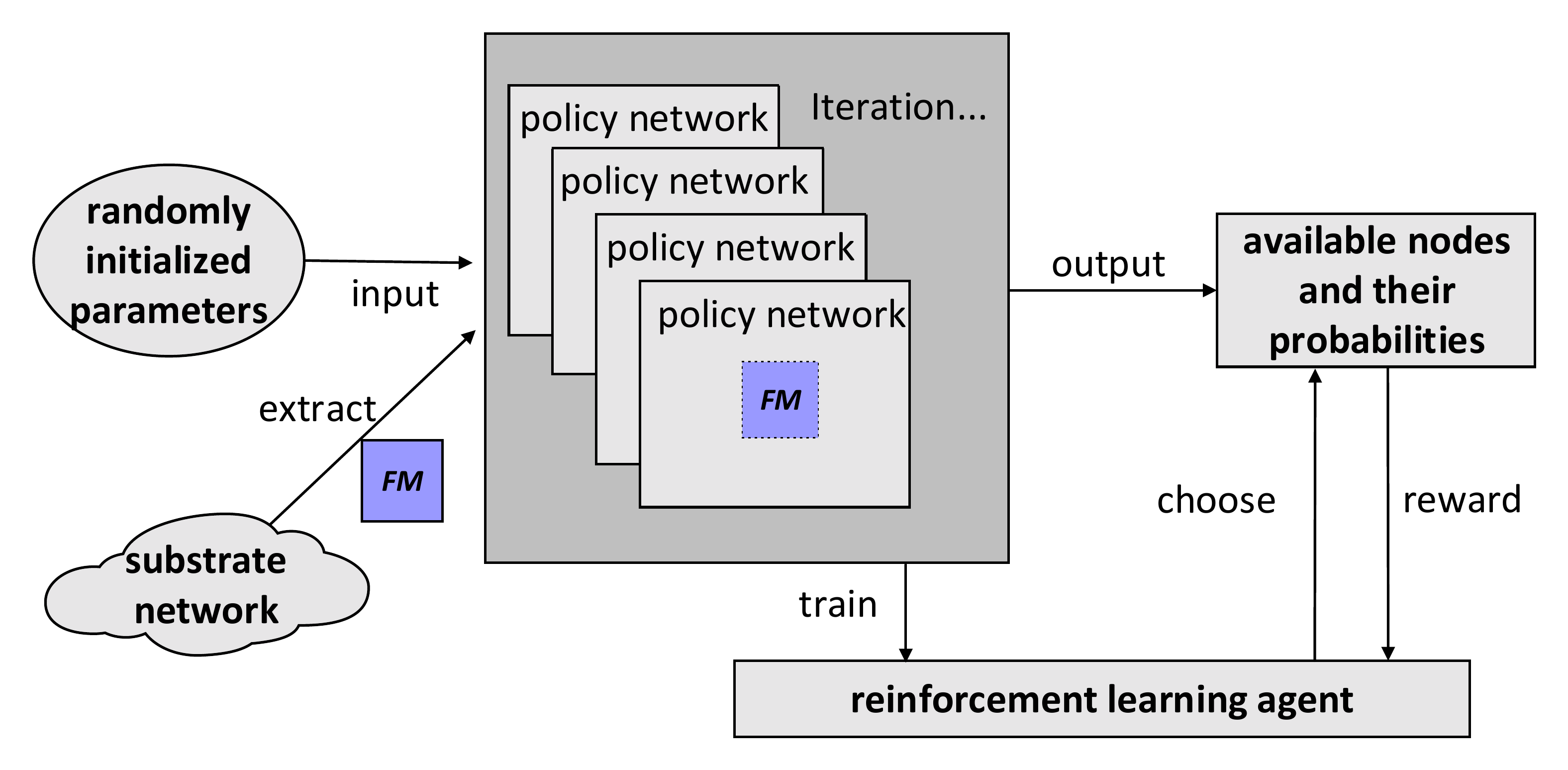}
\caption{Training process diagram.}
\label{fig_4}
\end{figure}

We use the policy network as a learning agent. First, the policy network is initialized to the unlearned state. After the feature matrix is input, we take the feature matrix as the learning environment of the agent. By fully learning each node attribute in the feature matrix, the agent selects those substrate nodes that satisfy both the computational resource requirements of the virtual node and the security performance requirements. The final policy network outputs a set of available substrate nodes and the probabilities that the virtual nodes map to them. After obtaining the probability of each substrate node, we use the probability distribution model to generate a sample from the substrate network set, from which a substrate node is selected as the node to be mapped. Because the initialization of the policy network is random, the node with the highest probability does not mean that it can be mapped to the optimal result. This process is repeated until all virtual nodes are allocated or VNE is terminated due to insufficient resources of the substrate node. If all virtual node mappings are successful, the link mapping continues. The training process can be represented as Figure \ref{fig_4}.

In RL, learning effect is determined by the action taken by learning agent, so we need to set a reward standard for learning agent. If the agent's current behavior can make the algorithm achieve greater benefits or better results, then the agent should be encouraged to continue to take the current action to obtain the cumulative reward. If the result of the agent's current action is small or harmful, the reward signal will become small or even disappear. The agent will stop the current action and take a new action instead. So an appropriate reward signal is very important. In the problem of VNE, we use the long-term revenue consumption ratio as a reward signal. This index reflects the utilization of the substrate resources, especially the link bandwidth resources. If the agent's current action can produce a higher revenue consumption ratio, then the agent will receive a larger reward signal, and continue to explore the action that produces a greater revenue consumption ratio. On the contrary, the agent stops its action and then takes a new action.

In the training process, we set a target symbol for each virtual node in the VNRs. This symbol represents the substrate node to which the virtual node is embedded. Assuming that the target symbol of virtual node $n_i^v$ is $j$, it means that the $j$-th dimension of the feature vector corresponding to substrate node $n_j^s$ is 1, and other dimensions are 0. It is expressed as follows:

\begin{equation}
\begin{aligned}
n_j^s=(0_1,0_2,...1_j...,0_k)^T.
\end{aligned}
\end{equation}

The next step is to output the error between the target vector $v_j^s$ and $n_j^s$, that is, the cross entropy loss:

\begin{equation}
\begin{aligned}
Loss(n_j^s,v_j^s)=-\sum_jn_j^slog(v_j^s).
\end{aligned}
\end{equation}

Then we use gradient descent algorithm to calculate the loss of gradient $g_f$:

\begin{equation}
\begin{aligned}
gl= g_f \cdot \alpha \cdot reward.
\end{aligned}
\end{equation}

Where $reward$ is the reward signal and $\alpha$ is the learning rate.

The learning rate $\alpha$ control calculates the size of the gradient. If the gradient is too large, the learning agent's action adjustment direction will be too large, may miss some of the more expensive action. Therefore, no amount of training can achieve better results. If the gradient is too small, the training of the agent will be extremely slow and waste a lot of time. Therefore, the learning rate should be carefully adjusted. We adopt batch gradient descent algorithm to update the strategy network, which not only improves the convergence speed of the agent training, but also guarantees the stability of the network \cite{B11}.

The training process is shown in algorithm 1.

\begin{algorithm}
  \caption{Training phase algorithm}
  \begin{algorithmic}[1]
    \Require
        {$number\,of\,epoches, epoch$; $learning\,rate, \alpha$; $trainingset$};
    \Ensure
        {$parameters\,of\,the\,policy\,network$};
    \While {$iteration<epoch$}
    \For {$request \in trainingset$}
    \State {$counter=0$};
    \For {$v\_node \in request$}
    \State $FM=getFeatureMatrix()$;
    \State $dis=getOutPut(FM)$;
    \If {$sn.cpu \geq vn.cpu\,and\,sl \geq sr$}
    \State $c=sample(dis)$;
    \State $getGradient(c)$;
    \EndIf
    \EndFor
    \If {$isMapped(\forall\,v\_node\!\in\!request)$}
    \State $BFSLinkMap(request)$; \quad
    \EndIf
    \If {$isMapped(\forall\,v\_node\!\in\!request)$}
    \If {$(\forall\,v\_link\!\in\!request)$}
    \State {$reward=revenue(request)$};
    \Else
    \State {$stacking\,gradient=0$};
    \EndIf
    \EndIf
    \State {$counter++$};
    \If {$counter == batch\_size$}
    \State {$counter = 0$};
    \EndIf
    \EndFor
    \State {$iteration++$};
    \EndWhile
    \State $return\,parameters$;
  \end{algorithmic}
\end{algorithm}

The input parameter epoch indicates that all training data will be sent to the policy network to complete a forward calculation and back propagation process. In each epoch, we input all VNRs for training. Line 5 of the algorithm indicates that the feature matrix is obtained. Line 6 is the probability distribution of the substrate nodes. Lines 7-10 represent the node mapping. Line 13 shows the link mapping. Lines 15-21 Compute the rewards obtained after the nodes and links are successfully embedded.

In the test phase, we select the node with the highest probability as the mapping node directly. The test process is shown in algorithm 2. Where line 5 represents the node map and line 8 represents the link map using BFS. Ends if all virtual nodes and virtual links are mapped successfully.

\begin{algorithm}
  \caption{Test phase algorithm}
  \begin{algorithmic}[1]
     \Require
        {$test\_set$};
    \Ensure
        {$long\,term\,average\,revenue$},
        {$acceptance\,rate$},
        {$long\,term\,revenue\,consumption\,ratio$};
    \State {$Initialize\,the\,policy\,network$};
    \For {$request \in test\_set$}
    \For {$v\_node \in request$}
    \State $FM=getFeatureMatrix()$;
    \State $candi\_node=getProbability(c)$;
    \EndFor
    \If {$isMapped(\forall\,v\_node\!\in\!request)$}
    \State $BFSLinkMap(request)$;
    \EndIf
    \If {$isMapped(\forall\,v\_node\!\in\!request, \forall\,v\_link\!\in\!request)$}
    \State $return\,(success)$;
    \EndIf
    \EndFor
  \end{algorithmic}
\end{algorithm}

\subsection{Algorithm Complexity Analysis}

The algorithm complexity of our proposed security-aware VNE algorithm based on RL is $O(C_{VNR}(C_{n^s} \cdot d+C_{n^v}+C_{l^v}))$. Where $C_{VNR}$ represents the number of incoming VNRs. $C_{n^s}$ represents the number of substrate nodes. $d$ is the dimension of the feature matrix. $C_{n^v}$ and $C_{l^v}$ represent the number of successfully embedded virtual nodes and the number of virtual links, respectively.

The specific derivation of algorithm complexity is shown in TABLE \uppercase\expandafter{\romannumeral2}.

\begin{table}
\centering
\caption{Algorithm complexity analysis}
\renewcommand\arraystretch{1.5}
\begin{tabular}{|p{50mm}|p{30mm}|}
\hline
Algorithm steps & Algorithm complexity  \\
\hline
complexity of computing feature matrix for every VNR & $O(C_{n^s} \cdot d)$ \\
\hline
the embedded complexity of every VNR computing node & $C_{n^v}$ \\
\hline
the embedded complexity of every VNR computing link & $C_{l^v}$ \\
\hline
complexity of successful embedding of every VNR & $O(C_{n^s} \cdot d+C_{n^v}+C_{l^v})$ \\
\hline
complexity of successful embedding of all VNRs & $O(C_{VNR}(C_{n^s} \cdot d+C_{n^v}+C_{l^v}))$ \\
\hline
\end{tabular}
\end{table}

\section{Experimental Setup and Result Analysis}

\subsection{Experimental Setup}
The substrate topology is generated by GT-ITM tool, which is commonly adopted in virtual network mapping algorithm \cite{B10,C3}, to form a substrate network with 100 nodes and 570 links. We set the computing resource and security level for each substrate node. The computing resource of each substrate node is evenly distributed between 50 and 100 units. The security level of each substrate node is evenly distributed between 0 and 3. In addition, we set the bandwidth resource of each substrate link to a range from 20 to 50.

Similarly, we generate 2000 VNRs, 1000 of which are used as training set and another 1000 as test set. Each of these requests has between 2 and 10 virtual nodes at random. The computing resource requirements of virtual nodes are uniformly distributed between 0 and 50 units. The security requirements of virtual nodes are evenly distributed between 0 and 3. The bandwidth resource requirements of the virtual link are evenly distributed between 0 and 50. Virtual nodes are connected with each other with a probability of 50\%. The arrival of virtual network request simulates the Poisson process. The summary of parameters is shown in TABLE \uppercase\expandafter{\romannumeral3}.

\begin{table}
\centering
\caption{Parameter setting}
\renewcommand\arraystretch{1.5}
\begin{tabular}{|p{40mm}|p{40mm}|}
\hline
Parameter names & Parameter values  \\
\hline
number of substrate nodes & 100 \\
\hline
number of substrate links & 570 \\
\hline
substrate node resource & U[50, 100] \\
\hline
substrate link resource & U[20, 50] \\
\hline
security level & U[0, 3] \\
\hline
number of nodes per VNR & U[2, 10]  \\
\hline
virtual node computing resource requirements & U[0, 50] \\
\hline
virtual link bandwidth resource requirements & U[0, 50] \\
\hline
virtual node connection probability & 50\% \\
\hline
safety requirements level & U[0, 3] \\
\hline
\end{tabular}
\end{table}

The experimental platform uses PyCharm and Python language to write experimental code. The training results and test results are shown in the diagram by Origin 8.5 . We used TensorFlow to build the policy network. TensorFlow is an open source software library for high-performance numerical calculations. With its flexible architecture, computing can be easily deployed to a variety of platforms \cite{A10}. Firstly, the four-tier structure of the policy network is constructed according to the description in section 4.2. The flexibility and ease of use of TensorFlow makes it easier to construct the four-tier structure. We initialize the policy network with parameters that conform to normal distribution. Set the learning rate of the learning agent to 0.005. We trained 100 epoch agents by gradient descent method.

\subsection{Training Results and Analysis}
\begin{figure}[!htp]
\includegraphics[width=1.0\columnwidth]{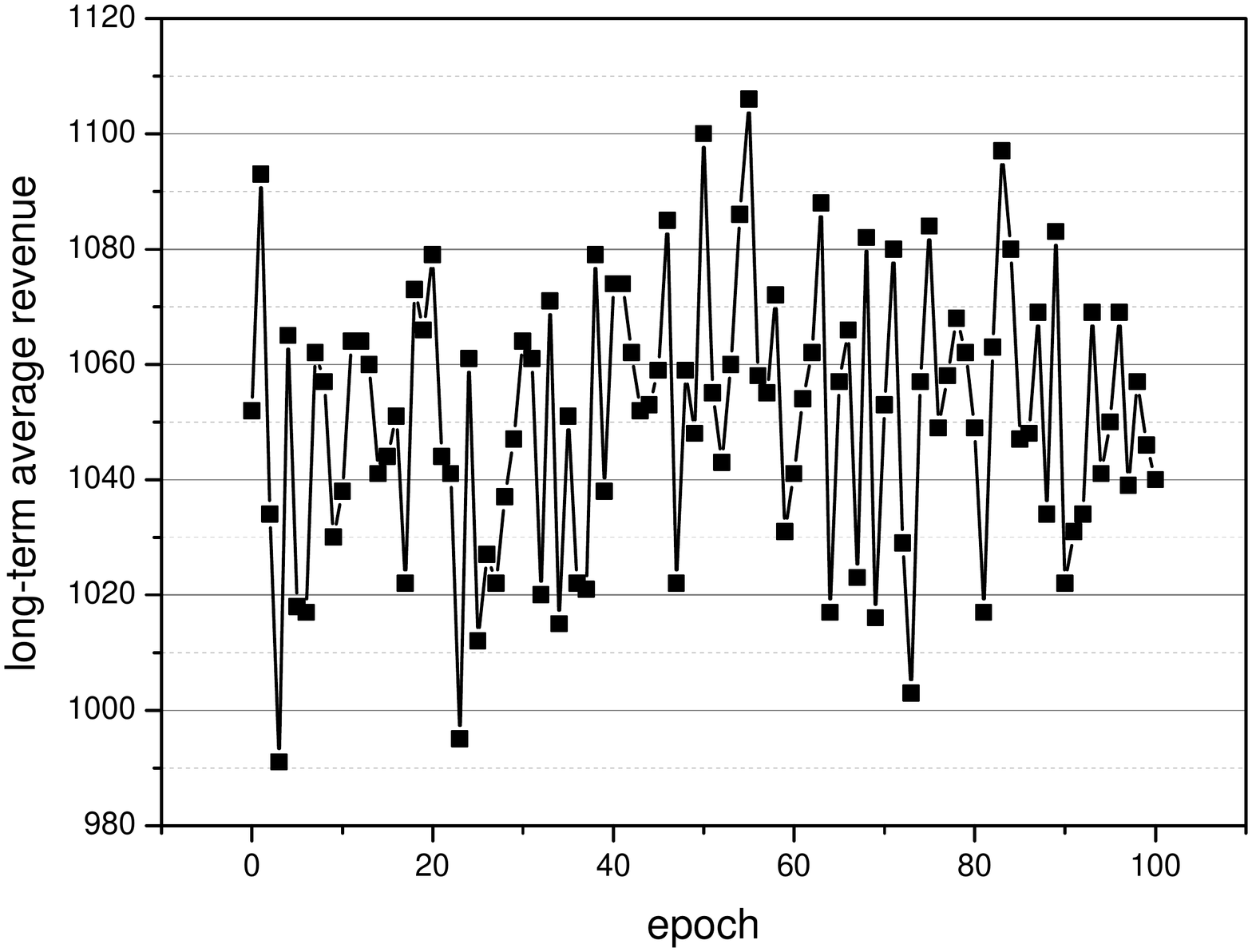}
\caption{Long-term average revenue on training.}
\label{fig_5}
\end{figure}

\begin{figure}[!htp]
\includegraphics[width=1.0\columnwidth]{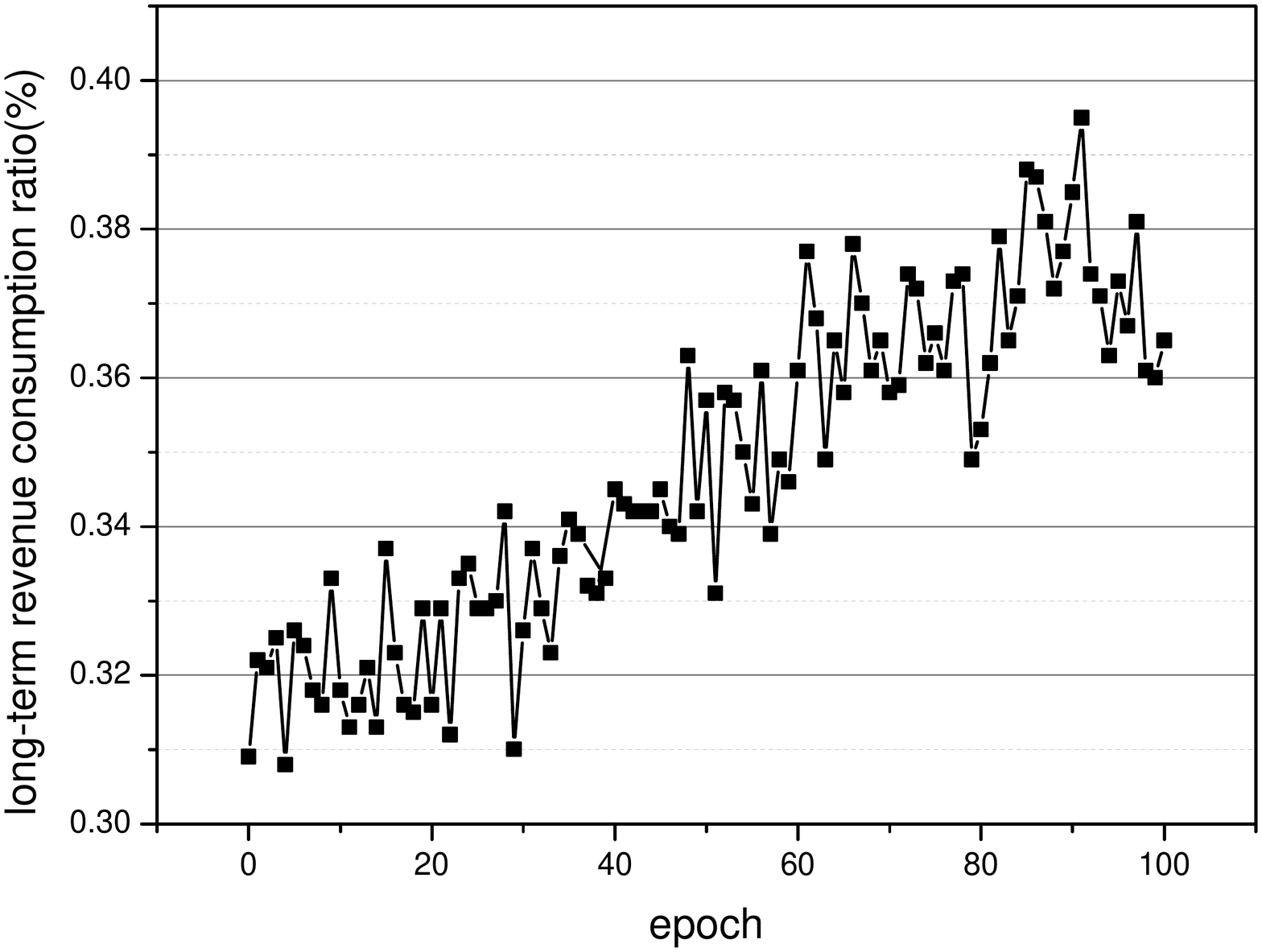}
\caption{Long-term revenue consumption ratio on training.}
\label{fig_6}
\end{figure}

\begin{figure}[!htp]
\includegraphics[width=1.0\columnwidth]{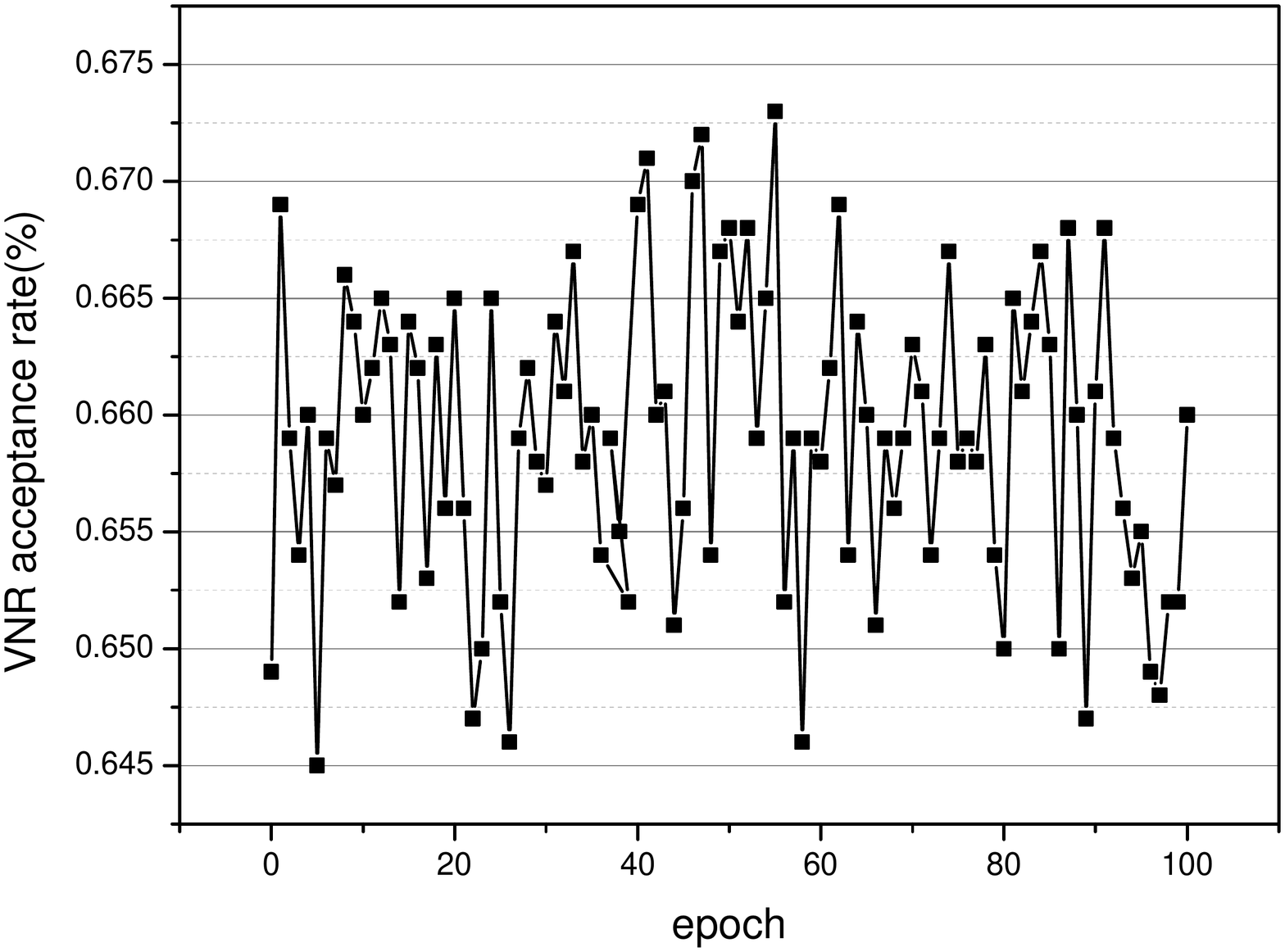}
\caption{VNR acceptance rate on training.}
\label{fig_7}
\end{figure}

Figure \ref{fig_5}, Figure \ref{fig_6} and Figure \ref{fig_7} show the changes in long-term average revenue, long-term revenue consumption ratio and VNR acceptance rate over 100 epochs, respectively.

As can be seen from the figures, the training process of RL is difficult to converge. Because RL agents need to constantly be aware of the state of the environment as they interact with it. We extract five attributes of the substrate node to form the feature matrix, which is used as the learning environment of the agent. The agent should fully learn how each feature might affect the final result. The agent is rewarded for making the decision. The training difficulty of RL agents, especially in NP-hard problems such as VNE, takes longer to converge.

In the early stage of training (0$<$epoch$<$30), the performance of the three evaluation indexes was very unstable. Because the parameters of the policy network are initialized randomly, the learning agent is completely unfamiliar with the current environment and can only take random actions, so the process does not converge. In the middle of training (30$<$epoch$<$80), the three evaluation indexes began to stabilize. In the process of training, agents begin to explore the substrate nodes with high probability. When choosing a substrate node with high probability, the agent will get rich rewards. So the agent will remember the benefits of this decision and make similar decisions in the future. For those agents who have not yet achieved better revenue, they will gradually choose to make their own actions more profitable. At the later stage of training (80$<$epoch$<$100), the three training indexes began to be more stable. This is because the model is still exploring substrate nodes with higher probability and most agents have been able to take actions to make their own profits. At this time, the training results can be considered as convergent.

\subsection{Test Results and Analysis}
\begin{figure}[!htp]
\includegraphics[width=1.0\columnwidth]{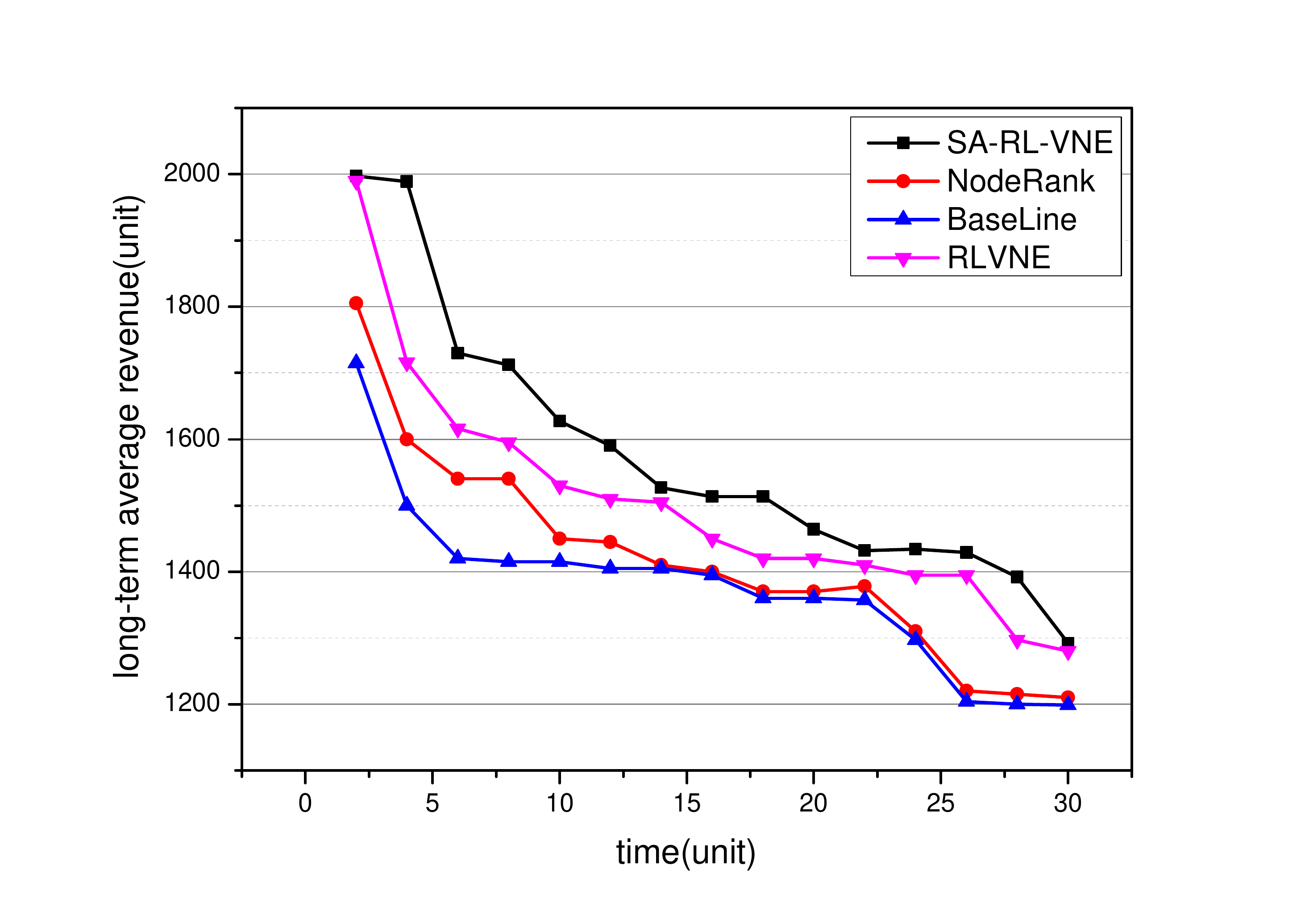}
\caption{Long-term average revenue.}
\label{fig_8}
\end{figure}

\begin{figure}[!htp]
\includegraphics[width=1.0\columnwidth]{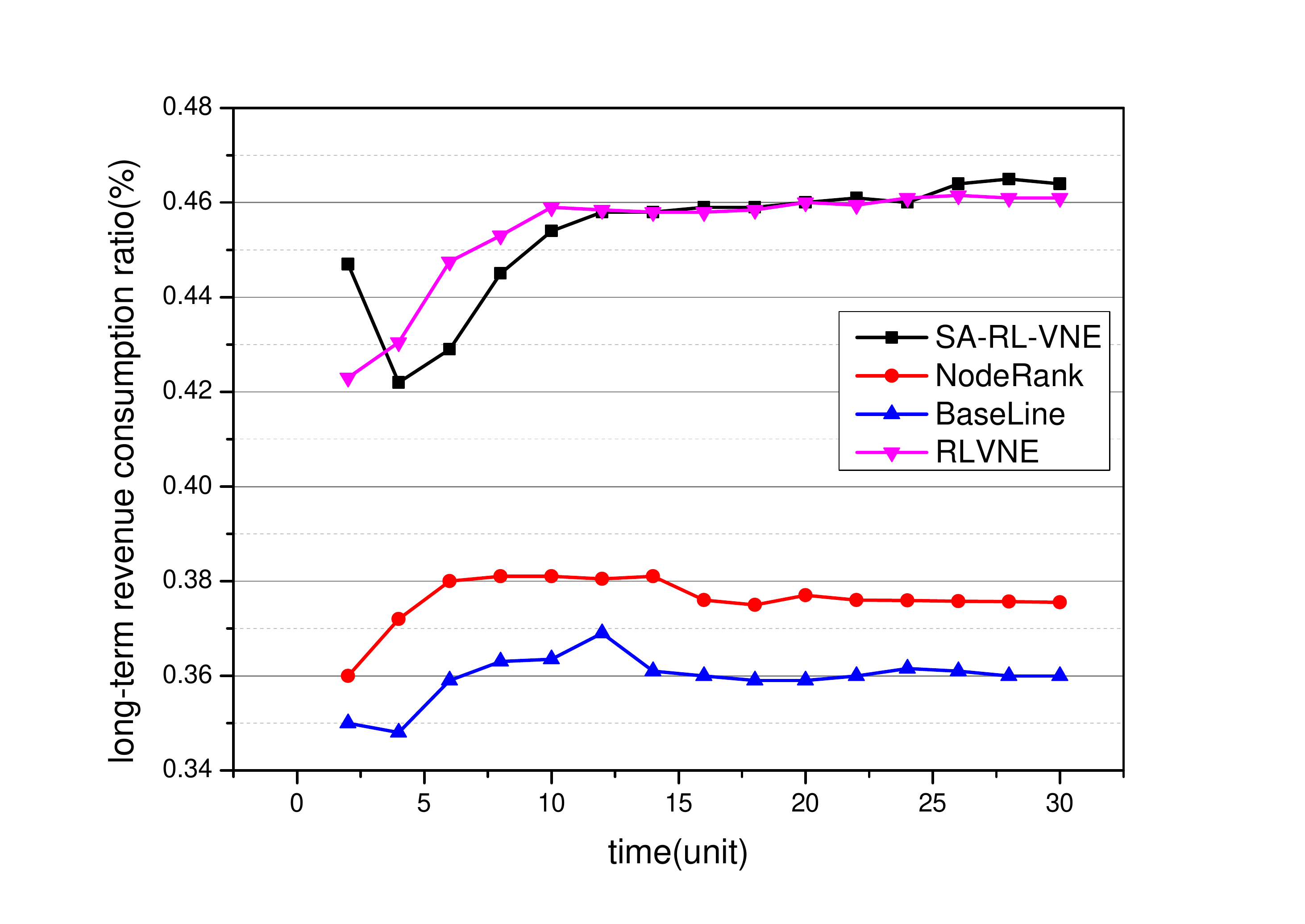}
\caption{Long-term revenue consumption ratio.}
\label{fig_9}
\end{figure}

\begin{figure}[!htp]
\includegraphics[width=1.0\columnwidth]{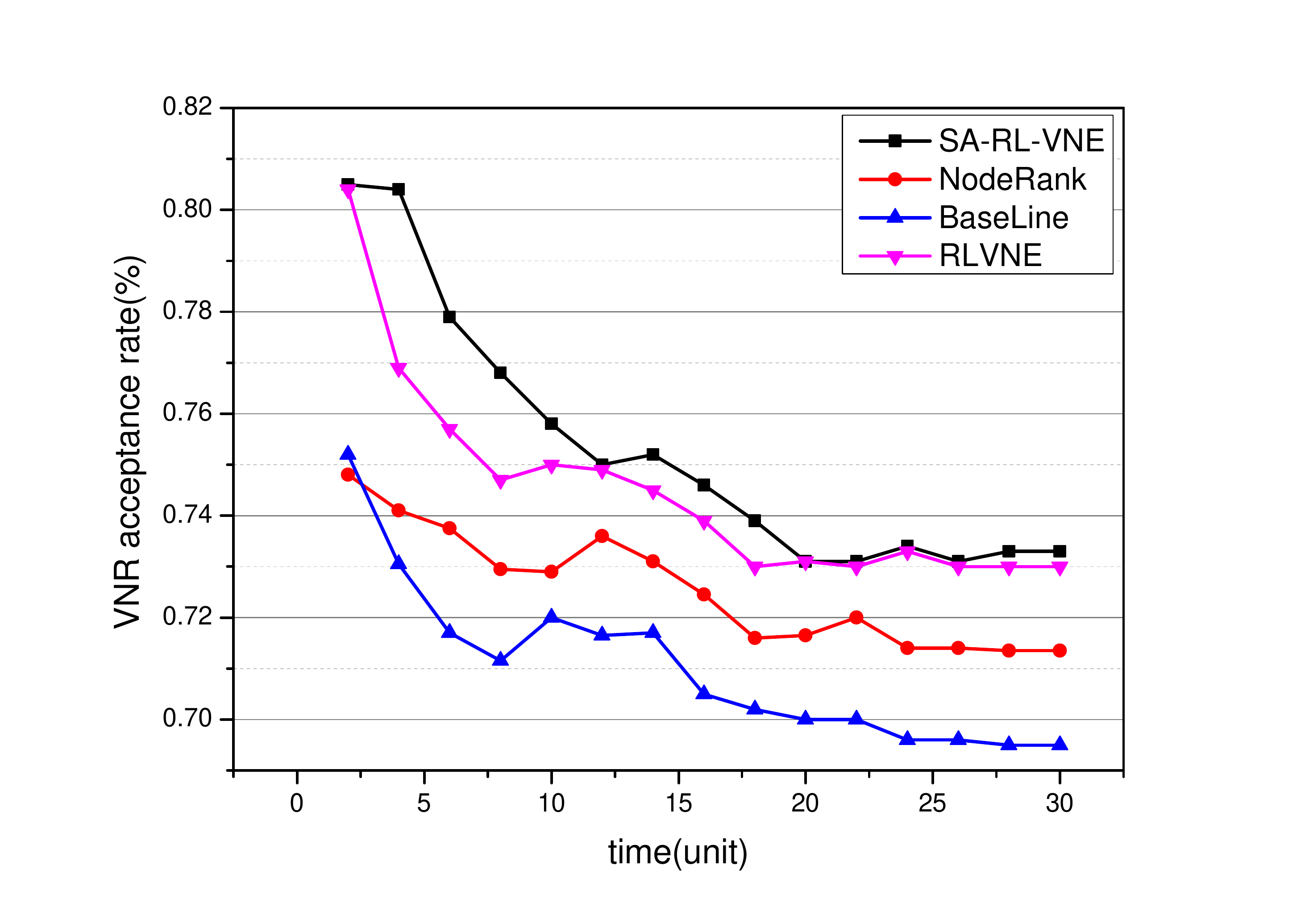}
\caption{VNR acceptance rate.}
\label{fig_10}
\end{figure}

In the test phase, we compared the security perception VNE algorithm based on RL (SA-RL-VNE) proposed in this paper with the RLVNE algorithm proposed in reference \cite{A9}, the BaseLine algorithm proposed in reference \cite{A5} and the NodeRank algorithm proposed in reference \cite{B9} in three aspects of long-term average revenue, long-term revenue consumption ratio and VNR acceptance rate. Because these three algorithms are based on BFS strategy to carry out link mapping, but they use different algorithms in the node mapping phase. RLVNE algorithm also trains the policy-network as a learning agent, but it does not pay attention to security factors. Similarly, the other two algorithms do not consider the security performance. Figure \ref{fig_8}, Figure \ref{fig_9} and  Figure \ref{fig_10} shows the test results on the test set.


It can be seen from Figure \ref{fig_8} that before 10 time units, the long-term average returns of the four algorithms are relatively high and decline rapidly. This is because in the early network of the substrate resources are relatively rich, almost all VNRs can be satisfied. The reason for the fast decline is that a large number of VNRs consume a large amount of resources and later VNRs will not all be satisfied. The index began to stabilize in the middle and late period. The situation is similar in Figure \ref{fig_10}. In the early stage, due to the rich substrate resources can meet more virtual network requirements, so the acceptance rate is high. With the continuous consumption of the substrate resources, the index began to decrease gradually in the later period. The SA-RL-VNE algorithm is superior to the other three algorithms in terms of long-term average return and VNR acceptance rate. The reason is that in the training phase, the agent of SA-RL-VNE algorithm can better learn the relationship between the substrate network nodes, which is more in line with the actual VNE situation. Therefore, the results on the test set are good.

It can be seen Figure \ref{fig_9} that the revenue consumption ratio is relatively stable in the whole process. Because this indicator has nothing to do with the number of substrate resources. This index depends on the efficiency of the algorithm. Generally speaking, the revenue consumption ratio of SA-RL-VNE algorithm is slightly higher than that of RLVNE. This is because the learning agent of SA-RL-VNE algorithm is trained in the environment with more characteristics of the substrate nodes, and has achieved better results. It shows that our algorithm not only meets the demand of VNR resources, but also considers the impact of security on VNR. Only the substrate nodes that meet the security level requirements can accept the embedding of virtual nodes, which improves the acceptance rate and ensures the security of the network.

In addition, we compare the average running time of these algorithms, as shown in TABLE 4. Because our policy network training needs a long time, plus the time required in the test phase. So the overall time is relatively long.

\begin{table}
\centering
\caption{The average running time of the algorithm}
\renewcommand\arraystretch{1.5}
\begin{tabular}{|p{30mm}|p{30mm}|}
\hline
The algorithm name & The run time(s) \\
\hline
SA-RL-VNE & 2780 \\
\hline
RLVNE & 2610 \\
\hline
NodeRank & 460 \\
\hline
BaseLine & 302 \\
\hline
\end{tabular}
\end{table}

\begin{figure}[!htp]
\includegraphics[width=1.0\columnwidth]{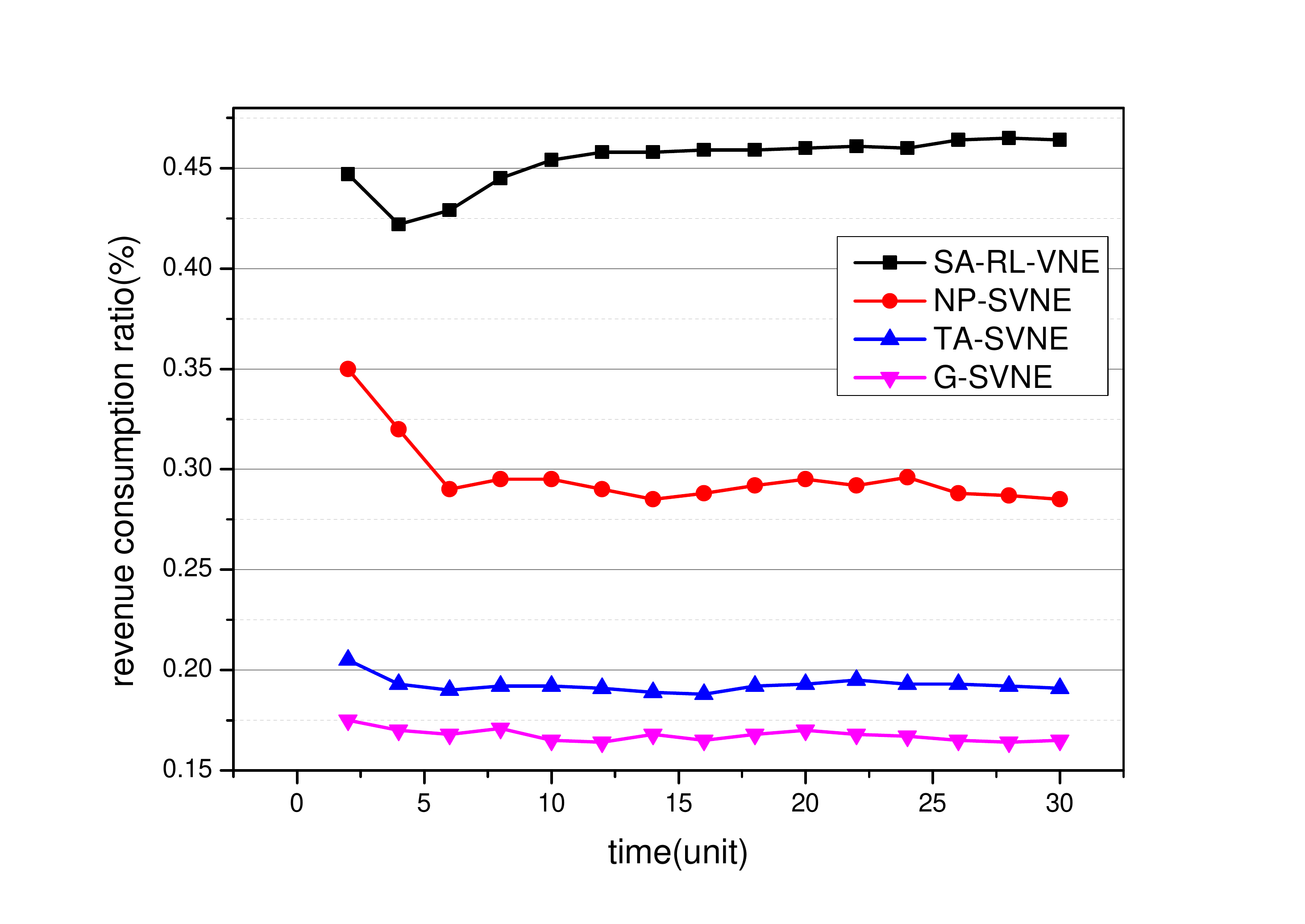}
\caption{Revenue consumption ratio.}
\label{fig_11}
\end{figure}

In order to highlight the SA-RL-VNE algorithm's efficiency is still better while considering security performance. We compare this algorithm with three typical VNE algorithms that consider security performance in terms of revenue consumption ratio. They are NP-SVNE algorithm proposed in \cite{A1}, TA-SVNE algorithm and G-SVNE algorithm in \cite{A2}. They are all based on heuristic secure VNE algorithms. The NP-SVNE algorithm sets the security attributes requested by the virtual network to different security requirement levels and security levels, instead of setting each virtual node to a different security requirement level and security level. The TA-SVNE algorithm and G-SVNE algorithm abstract the security attributes of virtual nodes and substrate nodes into security requirement levels and security levels. The algorithm established a security VNE mixed integer linear programming model. However, these two algorithms do not consider the impact of the node's security attributes on the embedding revenue consumption ratio. The comparison results are shown in Figure \ref{fig_11}.

The result shows the advantages of the intelligent learning algorithm. Because NP-SVNE is a heuristic algorithm, it cannot learn the characteristics of the substrate network through agents as intelligent algorithm. The overall efficiency of heuristic algorithm is not as high as intelligent algorithm. It can be seen that the SA-RL-VNE algorithm has better performance when security performance is considered, so using intelligent algorithm to solve the problem of VNE has important practical significance and will have a broader prospect.

\section{Conclusions and Future Work}
This paper presents a method of embedding security-aware virtual network based on RL. RL is mainly applied to virtual node embedding. Five attributes of the substrate node are extracted to form the feature matrix in order to provide a more real training environment for the RL agent. The VNE decision is entirely determined by the return to the agent, so there is no need to rely on any hand-written rules. This paper focuses on the security problem in the VNE. The virtual nodes can only be mapped to the substrate nodes that meet their security requirements. It is of great significance for network business with high security requirement in reality. Simulation results demonstrate that the performance of this algorithm is more preferable than other algorithms.

In the future work, we will explore to extract more reasonable features for each substrate node and create a more real network environment for the agent. In addition, we will further investigate how to train the agent better without using the feature matrix as input.

\section{Acknowledgement}

The authors would like to thank the editor and reviewers for their excellent work and constructive comments, which improved the paper quality. This work is partially supported by the Major Scientific and Technological Projects of CNPC (Grant No. ZD2019-183-006), and partially supported by ``the Fundamental Research Funds for the Central Universities" of China University of Petroleum (East China) (Grant No. 20CX05017A, 18CX02139A).

\ifCLASSOPTIONcaptionsoff
  \newpage
\fi

\end{document}